# How Intense Electromagnetic Wave Affects Photoassociation

M.A. Kutlan

Institute for Particle & Nuclear Physics, Budapest, Hungary

kutlanma@gmail.com

The influence of intense electromagnetic wave on photoassociation is investigated. The quantum-mechanical problem of the stimulated transition is considered. The probabilities for stimulated production of mesic atoms in states with large principal quantum numbers are calculated

1. **Introduction**

Transitions from the continuous spectrum to a bound state with spontaneous emission of a photon are possible in particle collisions. As a rule, however, the probability for such transitions is small. The probability for such transitions in atomic collisions may be considerably enhanced in the field of an intense electromagnetic wave, however, on account of stimulated photon emission [1-3].

The stimulated transitions that arise in collisions of ordinary atoms or ions were considered in the papers cited above. However, induced transitions in the field of an intense electromagnetic wave may also arise in the production of mesic atoms, muonium, positronium P [4-6], antiprotonium, mesic molecules, etc. In these cases the stimulated photoassociation process has a number of specific features that must be taken into account in calculating the corresponding probabilities.

In particular, in the production of the "exotic" atoms and molecules mentioned above at thermal energies, the wavelength of the colliding particles considerably exceeds the size of the system in the bound state. Moreover, electronic terms may be excited in collisions of ordinary atoms, but internal degrees of freedom are not excited in the production of exotic atoms (or molecules). This is similar to processes discussed in [10-18].

In this paper we consider the quantum-mechanical problem of the stimulated transition of a system from the continuous spectrum to a bound state that has a finite lifetime against spontaneous transitions to lower states (gamma-ray emission, the Auger effect, etc.). The resulting formulas will be used to calculate probabilities for stimulated production of mesic atoms in states with large principal quantum numbers and for the production of mesic molecule $dd\mu$ in collisions of mesic deuterium with deuterium in the field of an intense electromagnetic wave. The second process is possible in principle since the $dd\mu$ system has several bound states, including one with a binding energy of -2 eV [7].

2. **The reaction rate**

Let the energy of the particles before collision be $E_1 = \varepsilon_p + \varepsilon_1 + \varepsilon_2$, where $\varepsilon_p$ is the kinetic energy of the relative motion of the particles, and $\varepsilon_1$ and $\varepsilon_2$ are the internal energies of the colliding particles. We shall write the potential for the interaction of the system of particles with the external electromagnetic field in the dipole approximation:

$$V(t) = \tilde{\mathbf{d}} \mathbf{E}_0 \cos \omega t, \tag{1}$$

where $\omega$ and $\mathbf{E}_0$ are the frequency and electric field strength, respectively, of the external electromagnetic field and $\tilde{\mathbf{d}}$ is the operator for the transition dipole moment [6].

We shall seek the wave function of the system in the external electromagnetic field in the form ($\hbar = c = 1$)

$$\Psi(t) = A(t)\Phi_{E_1} + B_{\mathbf{p}_0\mathbf{n}}(t)\Phi_{E_n\mathbf{p}_0}^{*} + \sum_{\mathbf{p}'} D_{\mathbf{p}'}(t)\Phi_{E'} + \sum_{\lambda} C_{\lambda}\Phi_{\lambda}, \quad (2)$$

where $\Phi_{E_1}$ is the wave function of the system with energy $E_1$ in the continuous spectrum (the initial state), $\Phi_{E_n\mathbf{p}_0}^{*}$ the wave function of the bound quasistationary state of energy $E_n$ produced from the initial state of momentum $\mathbf{p}_0$, $\Phi_{E'}$ is the continuum wave function of the system of energy E' that results from the breakup of the bound quasistationary state $\Phi_{E_n\mathbf{p}_0}^{*}$, and $\Phi_{\lambda}$ is the wave function for the bound state to which the spontaneous transition from the state $\Phi_{E_n\mathbf{p}_0}^{*}$ goes. We shall use Heitler's method [8] to solve the time dependent Schriidinger equation with interaction (1). We subject the amplitudes $A(t)$, $B_{\mathbf{p}_0\mathbf{n}}$, $D_{\mathbf{p}'}(t)$, and $C_{\lambda}$ to Fourier transformations:

$$A(t) = -\frac{1}{2\pi i}\int dE\, A(E)\exp\{i(E_1 - E)t\},$$

$$B_{\mathbf{p}_0\mathbf{n}}(t) = -\frac{1}{2\pi i}\int dE\, B_{\mathbf{p}_0\mathbf{n}}(E)\exp\{i(E_{\mathbf{p}_0\mathbf{n}} - E)t\} \quad (3)$$

etc. After substituting (3) into the time dependent Schriodinger equation we obtain the following set of equations for the amplitudes ($\hbar = c = 1$):

$$(E - E_1 + i\varepsilon)A(E) = 1 + \sum_{\mathbf{p}_0} V_{\mathbf{p}_0 n} B_{\mathbf{p}_0 n}(E),$$

$$(E - E_{n\mathbf{p}_0} - \omega + i\varepsilon)B_{n\mathbf{p}_0}(E) = V_{\mathbf{p}_0 n}^{*} A(E) + \sum_{\mathbf{p}_0} V_{\mathbf{p}_0 \mathbf{p}} D_{\mathbf{p}}(E) + \sum_{\lambda} H_{\lambda} C_{\lambda}(E),$$

$$(E - E_{\lambda} - \omega + i\varepsilon)C_{\lambda}(E) = H_{\lambda}^{*} B_{n\mathbf{p}_0}(E), \quad (4)$$

$$(E - \varepsilon_{\mathbf{p}} + \varepsilon_{\mathbf{p}_0} + i\varepsilon)D_{\mathbf{p}}(E) = V_{\mathbf{p}_0 \mathbf{p}}^{*} B_{n\mathbf{p}_0}(E),$$

$$V_{\mathbf{p}_0 n} = \int \Phi_{E_n}^{*} \mathbf{d}(\mathbf{R})\mathbf{E}_0 \Phi_{E_1} d\mathbf{R},$$

where $H_{\lambda}$ is the matrix element for the transition from the bound state $\Phi_{E_n}^{*}$ to another bound state $\Phi_{\lambda}$ of the system with the emission, for example, of a photon, and $\lambda$ is the set of quantum numbers specifying the new bound state of the system and the emitted photon ($\mathbf{k}$ and $\mu$ are the wave vector and polarization of the emitted photon). Equations (4) were derived in the resonance approximation, i.e. under the assumption that $\omega \gg \varepsilon_{\mathbf{p}_0} + |I_0| - \omega$, where $I_0$ is the energy of the bound state reckoned from the energy of the atoms at infinite separation ($R \to \infty$). The solution of Eqs. (4) for the amplitude $C_{\lambda}(E)$ can be expressed in the form

$$C_\lambda(E) = H_\lambda V_{\mathbf{p}_0 n}[(E - E_\lambda - \omega + i\varepsilon)(E - E_{n\mathbf{p}_0} - \omega + i\Gamma/2)(E - E_1 + i\Gamma_0/2)]^{-1},$$

$$\Gamma_0 = 2\sum_{\mathbf{p}_0} \frac{|V_{\mathbf{p}_0 n}|^2}{E - E_{n\mathbf{p}_0} - \omega + i\Gamma/2}, \quad \gamma = 2\sum_\lambda |H_\lambda|^2 \zeta(E - E_1 + \omega), \quad (5)$$

$$\Gamma_{n\mathbf{p}} = 2\sum_{\mathbf{p}} |V_{\mathbf{p}_0 n}| \zeta(E - \varepsilon_\mathbf{p} - \varepsilon_{\mathbf{p}_0}), \quad \Gamma = \Gamma_{n\mathbf{p}} + \gamma,$$

where $\gamma$ is the width of the bound level due to the spontaneous decay with photon emission and $\Gamma_{n\mathbf{p}}$ is the width associated with the transition of the system from the bound state to the continuum under the action of the external electromagnetic field.

As is shown in Heitler's book [8] the transition probability per unit time to the state $\Phi_\lambda$ can be obtained with the aid of Eq. (5):

$$W = \frac{2\pi}{\hbar} \sum_\lambda \frac{|H_\lambda|^2 |V_{\mathbf{p}_0 n}|^2 \delta(E_1 - E_\lambda - \omega)}{(\varepsilon_{\mathbf{p}_0} + |I_0| - \omega)^2 + \Gamma^2/4}. \quad (6)$$

On summing (5) over $\lambda$ and dividing (6) by the particle flux, we obtain the cross section for the process at a fixed value of energy $\varepsilon_{\mathbf{p}_0}$:

$$\sigma = g\pi \frac{\hbar^2}{\mathbf{p}_0^2} \frac{\Gamma_{n\mathbf{p}} \gamma}{(\varepsilon_{\mathbf{p}_0} + |I_0| - \omega)^2 + \Gamma^2/4}, \quad (7)$$

where g is the statistical weight and $\mathbf{p}_0$ is the momentum of the particles. We shall obtain expressions for $\Gamma_{n\mathbf{p}}$ for two cases: a) systems with Coulomb interaction (mesic atoms, muonium, positronium, antiprotonium, etc.) and b) the $dd\mu$ system. For the systems listed under case a), resonant transitions in the optical range are possible only to excited levels with $n \gg 1$. Hence to evaluate $\Gamma_{n\mathbf{p}}$ we may use the asymptotic expression for the radial wave function of the quasistationary level (in atomic units of length)

$$\Phi_{n\mathbf{p}_0}^* \approx 2^{n+1/2} r^{n-1} e^{-r/n} / (2n!)^{1/2} n^{n+1/2}. \quad (8)$$

As a result we obtain the following estimate for $\Gamma_{n\mathbf{p}}$ (for $n \gg 1$ and $kna_B \ll 1$):

$$\hbar\Gamma_{n\mathbf{p}} \approx \frac{2}{3} \frac{1}{\sqrt{\pi}} n^{19/2} (eE_0 a_B)^2 a_B^3 |\mathbf{p}_0| m/\hbar^3, \quad (9)$$

where $a_B$ is the Bohr radius. In deriving (9) it was assumed that the transition to the bound state takes place from an **S** state of the continuum.

For mesic atoms, binding energies in the optical range correspond to $n \gg 30$. In this case we obtain

$$\hbar\Gamma \sim 10^{-13} (eE_0)^2 (kT)^{1/2}. \text{ eV} \quad (10)$$

for thermalized muons. The reaction rate $\lambda$, averaged over the Maxwell distribution of the colliding particles, is given by

$$\hbar\lambda = \hbar n_0 \int v d\sigma = \left(\frac{2}{\pi}\right)^{1/2} n_0 \frac{\gamma}{kT} \frac{|V_{n\mathbf{p}}|^2}{kT} \int_0^\infty x^{1/2} e^{-x} [(x-\Delta)^2 + b^2]^{-1} dx, \qquad (11)$$

where $n_0$ is the particle density, $x = \varepsilon_{\mathbf{p}_0}/kT$, $\Delta = (\omega - |I_0|)/kT$, and $b = \Gamma/kT$. For $\Delta \gg 1$ and $b \ll \Delta$, we obtain

$$\hbar\lambda \approx 2(2\pi)^{1/2} n_0 \frac{|V_{n\mathbf{p}}|^2}{kT} \Delta^{1/2} e^{-\Delta} \frac{\gamma}{\Gamma} \text{ eV} \qquad (12)$$

from Eq. (11), while for $b \gg 1$ and $\Delta \leq 1$, we obtain

$$\hbar\lambda \approx n_0 \frac{|V_{n\mathbf{p}}|^2}{\Gamma} \frac{\gamma}{\Gamma} \text{ eV}. \qquad (13)$$

Thus, the reaction rate for the production of bound systems with Coulomb interaction in the field of an intense electromagnetic wave ceases to be temperature dependent when the fieldstrength is high enough.

Let us obtain numerical estimates of the reaction rate for the production of mesic hydrogen atoms. Using (12), we find

$$\hbar\lambda \approx 10^{-30} n^{19/2} \frac{e^2 E_0^2 |V_{n\mathbf{p}}|^2}{kT} \frac{\gamma}{\Gamma} n_0 \qquad (14)$$

At the temperature and density of liquid hydrogen we have $kT \sim 10^{-3}$ eV and $n_0 = 4.25 \times 10^{22} cm^{-3}$; then assuming an electromagnetic field strength of $E_0 \sim 3 \times 10^4$ V/cm and using the values $n \sim 30$ and $\gamma \sim 10^{-4}$, we obtain $\hbar\lambda \sim 10^{-5}$ eV. This estimate shows that the time for spontaneous production at very moderate strengths of the electromagnetic field.

Now let us consider the production of mesic deuterium molecules in the field of an intense electromagnetic wave. In this case we must find the matrix element $V_{n\mathbf{p}}$ for the transition from the continuous spectrum of the $a + d\mu$ system to a weakly bound state of the $dd\mu$ [9]. The cross section for this process is given by Eq. (7), and the reaction rate, averaged over the Maxwell distribution, is given by Eq. (11), in which the matrix element is taken from [7]. According to [7], the level of the mesic molecule $dd\mu$ with quantum numbers J = 1 and v = 1 has the binding energy $|I_0| = 2.196$ eV and the corresponding matrix element is $V_{n\mathbf{p}} = 400$ mesic atomic units. We accordingly obtain the following estimate for the stimulated photoassociation width $\Gamma_{n\mathbf{p}}$ (at liquid-hydrogen density):

$$\hbar\Gamma_{np} \approx 10^{-20} (eE_0)^2 (kT)^{1/2} \text{ eV}. \qquad (15)$$

If $\Gamma_{n\mathbf{p}} \leq \gamma$, then $\gamma/\Gamma \sim 1$, and Eq. (12) therefore gives the following estimate of $\lambda$ (using $\Delta \sim 1$ and $n_0 = 4.25 \times 10^{22} cm^{-3}$):

$$\hbar\lambda \approx 10^{-25} (eE_0)^2 (kT) \text{ eV}. \qquad (16)$$

At liquid hydrogen temperature with an electric field of strength $E \sim 10^4$ V/cm, (16) yields $\lambda \sim 10^{-9}$ eV or $\lambda \sim 10^6 \sec^{-1}$. This is an order of magnitude larger than the measured probability for spontaneous production of mesic molecules in liquid hydrogen [7].

### 3. Conclusion

The production in the field of an intense electromagnetic wave of bound states in positronium and muonium atoms, mesic atoms, and mesic molecules at normal temperatures takes place from an S state of the continuum. Hence the formulas derived above are not directly applicable to these cases. Several partial waves of the continuum wave function may take part in the production of molecules in collisions of ordinary atoms. This case therefore requires some refinement.

We also note that the case of elastic scattering in the field of an electromagnetic wave requires special treatment. Thus, the negative level leads to the appearance in the elastic scattering amplitude of a pole term that may interfere with the potential-scattering amplitude. Interference in the inelastic channel may be neglected if the condition $eE_0 v / \hbar \omega^2 \ll 1$ is satisfied.

The above estimates of the probabilities for the production of mesic atoms in the field of an intense electromagnetic wave are valid provided the muons are thermalized. If the capture of a muon to form a mesic atom as a result of the Auger effect on bound electrons takes place from continuum states with energies $E_k \gg 1$ eV and the muon has not been thermalized, then one will not be able to significantly alter the probability for the production of mesic atoms by the use of optical lasers. But if the muons are thermalized, one can considerably affect the production of mesic atoms (and of other exotic atoms) by the use of lasers, and in particular, it becomes possible to control the cascade of mesic X rays and the populations of the states of the mesic atoms.